\documentclass[10pt,draft]{dis03}
\usepackage{epsf,amsmath}

\begin{document}

\title{NEW IDEAS FOR TRANSVERSE SPIN ASYMMETRIES IN INCLUSIVE REACTIONS  
\thanks{ Talk presented at DIS2003, XI International Workshop on Deep Inelastic Scattering,
St. Petersburg, 23-27 April,2003}}

\author{JACQUES SOFFER \\
Centre de Physique Th\'eorique, CNRS Luminy Case 907, \\
13288 Marseille Cedex 09, France \\
E-mail: soffer@cpt.univ-mrs.fr }

\maketitle

\begin{abstract}
\noindent We will give convincing arguments to view transverse spin 
asymmetries in inclusive reactions, as very exciting observables,
although there were considered as irrelevant for a long time.
We look forward to new high energy data in the near future.
\end{abstract}

\section{Motivation} 
Let us consider the inclusive reaction
$pp \to c X$, with the center-of-mass (c.m.) energy $\sqrt{s}$ and where
$c = \pi, \gamma, \Lambda, \mbox {jet, etc} $, with Feynman variable $x_F$ and
transverse momentum $p_T$. 
We will essentially discuss the relevance of two observables which
can be measured in this reaction, when the initial protons are
transversely polarized, namely the single-spin asymmetry (SSA), defined as
\begin{equation}
A_N^{c}(\sqrt{s}, x_F,p_T) =\frac{ d\sigma^{\uparrow}_{c} - d\sigma^{\downarrow}_{c}}
{d\sigma^{\uparrow}_{c} + d\sigma^{\downarrow}_{c}}
\end{equation}
and the double-spin asymmetry (DSA), defined as
\begin{equation}
A_{NN}^{c}(\sqrt{s}, x_F,p_T) =\frac{ d\sigma^{\uparrow \uparrow}_{c} - 
d\sigma^{\uparrow \downarrow}_{c}}
{d\sigma^{\uparrow \uparrow}_{c} + d\sigma^{\uparrow \downarrow}_{c}}~,
\end{equation}
sometimes also denoted by $A_{TT}$. Clearly, it is legitimate to ask why these
spin observables are important at high energy and what is the appropriate
kinematic region $x_F,p_T$, where they should be best investigated, both theoretically
and experimentally. For the SSA in $pp \to \pi X$, there are interesting
data at $\sqrt{s} = 19.4 \mbox {GeV}$ from FNAL \cite{E704} and 
for $\pi^{0}$ production at $\sqrt{s} = 200 \mbox {GeV}$ from BNL-RHIC \cite{STAR}.
We will briefly recall the QCD mechanisms which have been proposed to explain
these data. We will also consider the case of the $W^{\pm}$ production
and we will show the usefulness of positivity for spin observables.

\section{Leading-twist QCD mechanisms for SSA and $W^{\pm}$ production}
By using the generalized optical theorem, one can write
\begin{equation}
A_{N}^{c}d\sigma=Im[f^*_{+}f_{-}]~,
\end{equation}
where $d\sigma = d\sigma_{c}^{\uparrow} + d\sigma_{c}^{\downarrow}$ is the corresponding 
unpolarized inclusive cross section. It is described by means of $f_{+}$, 
the forward {{\it non-flip} 
$3\rightarrow3$ helicity amplitude $ab\bar{c}_{\lambda} \rightarrow ab\bar{c}_{\lambda}$, 
where $\lambda=\pm $ is the same on both sides. Moreover $f_-$ is the forward 
{\it flip} amplitude $ab\bar{c}_{\lambda} \rightarrow ab\bar{c}_{-\lambda}$. 
In order to get a non-vanishing  $A_N^{c}$, one needs, a non-zero $f_-$ and furthermore 
it should have a phase difference with $f_+$. This point is important and has to be taken 
seriously, if we want to have a real understanding of the available experimental data. 
It is another way to say that a non-zero $A_{N}^c$ corresponds to a non-trivial situation, which 
reflects a high coherence effect among many different inelastic channels. However
according to naive parton model arguments one expects,
$A_N^c = 0$, but several possible mechanisms have been proposed recently to 
generate a non-zero $A_N^c$. They are based on the introduction of a transverse 
momentum (${\bf k}_T$) dependence of either the distribution functions $q(x,{\bf k}_T)$, 
for the Sivers effect \cite{DS} or of the fragmentation function $D_q(z,{\bf k}_T)$, for the 
Collins effect \cite{JCC}. These leading-twist QCD mechanisms have been used 
for a phenomenological study of this SSA \cite{AM} and higher-twist effects
have been also considered \cite{QS,KK}. Once we have identified these two fundamental 
leading-twist QCD mechanisms to generate SSA, the Sivers and Collins effects, in
order to study them in more detail it is important to be able to
discriminate between them. One way to achieve
this is to consider weak interaction processes, as proposed in
Ref. \cite{BHS3}. We recall that with the Collins mechanism, the
SSA is obtained from the transversity distribution function
$h_1^q$ of a quark of the initial polarized hadron, convoluted
with the Collins, ${\bf k}_T$ - dependent fragmentation function.
However in weak interaction processes, such as neutrino DIS on a
polarized target or $W$ production in polarized hadron-hadron
collisions, since the charged current only couples to quarks of
one chirality, $h_1^q$ decouples and thus the observed SSA is not
due to the Collins effect. This is not the case for the Sivers
effect which will be able to generate a non-zero SSA, as we will
see now.

Let us consider the inclusive production of a $W^{+}$ gauge boson in the
reaction $pp^{\uparrow}\rightarrow W^{+}X$, where one proton beam
is transversely polarized. In the Drell-Yan picture in terms of the
dominant quark-antiquark fusion reaction, the unpolarized cross-section reads
\begin{equation}
d\sigma=\int dx_ad^2{\bf k}_{Ta}dx_bd^2{\bf k}_{Tb}
[u(x_a,{\bf k}_{Ta})\bar d(x_b,{\bf k}_{Tb)}+ (u \leftrightarrow \bar d)]
d\hat \sigma^{ab\rightarrow W^{+}}.
\end{equation}
Similarly, the SSA can be expressed such as
\begin{eqnarray}
d\sigma A_N^{W^+} &=& \int dx_ad^2{\bf k}_{Ta}dx_bd^2{\bf k}_{Tb}
[\Delta^{N}u(x_a,{\bf k}_{Ta})\bar d(x_b,{\bf k}_{Tb})
\nonumber \\
&& -\Delta^{N}\bar d(x_a,{\bf k}_{Ta})u(x_b,{\bf k}_{Tb}) ]
d\hat \sigma^{ab\rightarrow W^{+}}.
\end{eqnarray}
For the Sivers functions \cite {DS} we have
\begin{eqnarray}
\Delta q^N(x,{\bf k}_T)&=&
q_{\uparrow}(x,{\bf k}_T) -  q_{\downarrow}(x,{\bf k}_T)
\nonumber \\
&& = q_{\uparrow}(x,{\bf k}_T) -  q_{\uparrow}(x,-{\bf k}_T)=
\Delta q^N(x,k_T){\bf S}_p\cdot{\bf \hat{p}}\times {\bf k}_T.
\end{eqnarray}
Here ${\bf S}_p$ denotes the transverse polarization of the
proton of three-momentum ${\bf p}$ and ${\bf \hat {p}}$ is a unit vector 
in the direction of ${\bf p}$. A priori the ${\bf k}_T$
-dependence of all these parton distributions is unknown, but as a
first approximation one can assume a simple factorized form for
the distribution functions and take for example, 
\begin{equation}
q(x,k_T)= q(x)f(k_T),
\end{equation}
where $f(k_T)$ is flavor independent, and a similar expression for
the corresponding Sivers functions. In such a situation, it is
clear that the SSA will also factorize and then it reads
\begin{equation}
A_N^{W^+} (\sqrt{s},y,{\bf p}_T)= H(p_T)A^{+}(\sqrt{s},y){\bf S}_p
\cdot{\bf \hat{p}}\times {\bf {p}}_T,
\end{equation}
where ${\bf p}_T$ is the transverse momentum of the $W^+$
produced at the c.m. energy $\sqrt{s}$ and $H(p_T)$ is a function
of $p_T$, the magnitude of ${\bf p}_T$. Obviously if the outgoing
$W^+$ has no transverse momentum, the SSA will be zero , as
expected. In the above expression we have now
\begin{equation}
A^{+}(\sqrt{s},y)=\frac{\Delta^{N}u(x_a)\bar d(x_b) - \Delta^{N}\bar d(x_a)u(x_b)}
{u(x_a)\bar d(x_b) + \bar d(x_a)u(x_b)},
\end{equation}
where $y$ denotes the $W^+$ rapidity, which is related to $x_a$
and $x_b$. Actually we have $x_a =\sqrt{\tau} e^y$ and  $x_b
=\sqrt{\tau} e^{-y}$, with $\tau =M_W^{2}/s$, and we note that a
similar expression for $A_N^{W^-}$, the SSA corresponding to $W^-$
production, is obtained by permuting $u$ and $d$. For the
$y$-dependent part of the SSA, one gets for $y=0$
\begin{equation}
A^{+}=\frac{1}{2}(\frac{\Delta^{N}u}{u} - \frac{\Delta^{N}\bar d}{\bar d}) 
 \  \  \  \ \mbox{and} \  \  \ \
A^{-}=\frac{1}{2}(\frac{\Delta^{N}d}{d} - \frac{\Delta^{N}\bar u}{\bar u})
\end{equation}
evaluated at $x = M_W /\sqrt{s}$.
Moreover a real flavor separation can be obtained away from $y=0$, since 
for $y=-1$ one has
\begin{equation}
A^{+}\sim - \frac{\Delta^{N}\bar d}{\bar d}  \  \   \  \ \mbox{and}  \  \  \  \
A^{-}\sim - \frac{\Delta^{N}\bar u}{\bar u}
\end{equation}
evaluated at $x=0.059$ and for $y=+1$ one has
\begin{equation}
A^{+}\sim  \frac{\Delta^{N}u}{u} \  \   \  \ \mbox{and}  \  \  \  \
A^{-}\sim  \frac{\Delta^{N}d}{d}
\end{equation}
evaluated at $x=0.435$, at a c.m. energy  $\sqrt{s}=500
\mbox{GeV}$. So the region $y\sim -1$ is very sensitive to the
antiquark Sivers functions, whereas the region $y\sim +1$ is sensitive
to the quark Sivers functions. As well known, the parity violating-asymmetry
$A_L^{W^{\pm}}$ allows the flavor separation of the quark helicity
distributions \cite{BS93}, similarly the measurement of $A_N^{W^{\pm}}$ is a
practical way to separate the $u$ and $d$ quarks Sivers functions
and their corresponding antiquarks $\bar u$ and $\bar d$. A
straightforward interpretation of a non-zero $A_L^{W^{\pm}}$ is,
in fact, a little bit more complicated because of its $p_T$ -
dependence, namely the factor $H(p_T)$ in Eq. (8), which is
unknown. It is possible to avoid this difficulty and to increase
statistics by integrating over the $p_T$ - range of the produced
$W^{\pm}$'s. This is part of the reason why we cannot make any
reliable prediction for these SSA, but the observation of
significant effects will be the unambiguous signature for the
presence of non-zero Sivers functions. The connection with
the spin physics programme at BNL-RHIC and further considerations are
also mentioned in Ref. \cite{ss13}.

\section{What positivity can bring into this game?}

The relevance of {\it positivity} in spin physics, which
puts non-trivial model independent constraints on spin observables, has been largely
discussed in Ref. \cite{JSa}.
These positivity conditions are based on the positivity
properties of density matrix or Schwarz inequalities for transition matrix elements in processes 
involving several particles carrying a non-zero spin. This important point was
emphasized by means of several examples chosen in different areas of particle physics, in
particular, total cross sections in pure spin states, two-body exclusive reactions,
polarized deep inelastic scattering, quark transversity distributions, off-shell 
gluon distributions, transverse momentum dependent distributions, single-particle 
inclusive reactions, polarized fragmentation functions, off-forward parton distributions, etc..
Very recently \cite{JSb}, new general positivity bounds were obtained, among the spin
observables in a single particle inclusive reaction, where the two initial particles
carry a spin-1/2. One of the consequences of this result is that, at rapidity
$y = 0$, one has
\begin{equation}
 1 + A_{NN} \geq 2|A_N|~,
 \end{equation}
for any value of $\sqrt{s}$ and $p_T$. In the case of $pp$ collisions, for some specific
reactions like $pp \to \mbox{jet} X$ or $pp \to \gamma X$, an estimate of $A_{NN}$ for $y \sim 0$
was done \cite{SSV} and it was found using Ref. \cite{JS}, that $A_{NN} \sim 0$. As a 
result one gets a much stronger bound on the corresponding SSA, namely $|A_{N}| \leq 1/2$.
This new non-trivial constraint will be very useful for model builders and
to check future data on SSA.

\section*{Acknowledgements} The author is thankful to INTAS (Project
587, call 2000), which provided the financial support for his attendance
to DIS2003.

\end{document}